\documentclass[twocolumn, singlespacing, tighten]{aastex631}

\usepackage{xspace,amsmath,amssymb}
\usepackage{mathptmx,txfonts,tikz,bm} 
\usepackage{xcolor, fontawesome}
\usepackage{tabularx}
\usepackage[LGR,T1]{fontenc}

\usepackage[utf8]{inputenc}
\usepackage{textcomp} 
\definecolor{linkcolor}{rgb}{0.1216,0.4667,0.7059}
\definecolor{xlinkcolor}{cmyk}{1,1,0,0}

\newcommand{\numax}{\mbox{$\nu_{\rm max}$}}
\newcommand{\Teff}{\mbox{$T_{\rm eff}$}}

\newcommand{\kepler}{{\em Kepler}\xspace}

\newcommand{\dnu}{\mbox{$\Delta{\nu}$}}

\newcommand{\logg}{\ensuremath{\mathrm{log}\ g}}

\newcommand{\tess}{\mbox{\textit{TESS}}\xspace}

\newcommand{\feh}{\ensuremath{[\mathrm{Fe/H}]}\xspace}
\newcommand{\afe}{\ensuremath{[\mathrm{\alpha/Fe}]}\xspace}

\usepackage{booktabs}
\usepackage{CJKutf8}

\shorttitle{Asteroseismology of Gaia BH2/3}
\shortauthors{Hey, Ong, and Li}

\graphicspath{{./}{figures/}}
\newcommand{\pysyd}{\texttt{PySYD}\xspace}
\newcommand{\pbjam}{\texttt{PBJam}\xspace}

\newcommand{\bha}{\mbox{BH2$^{*}$}\xspace}
\newcommand{\bhb}{\mbox{BH3$^{*}$}\xspace}

\newcommand{\bhadnu}{\ensuremath{5.99 \pm 0.03\ \mathrm{\mu Hz}}}
\newcommand{\bhanumax}{\ensuremath{60.15 \pm 0.57\ \mathrm{\mu Hz}}}
\newcommand{\bhaage}{\ensuremath{5.03^{+2.58}_{-3.05}\ \mathrm{Gyr}}}
\newcommand{\bharot}{\ensuremath{398 \pm 5\ \mathrm{d}}}

\usepackage{xcolor}
\newcommand{\rev}[1]{\textcolor{black}{#1}}

\begin{document}
\begin{CJK*}{UTF8}{gbsn}

\title{Asteroseismology of the red giant companions to Gaia BH2 and BH3}

\correspondingauthor{Daniel Hey}
\email{dhey@hawaii.edu}
\author[0000-0003-3244-5357]{Daniel Hey}
\altaffiliation{Not a Fellow}
\affiliation{Institute for Astronomy, 
    University of Hawaii,
    Honolulu, USA}

\author[0000-0003-3020-4437]{Yaguang Li (李亚光)}
\altaffiliation{Beatrice Watson Parrent Fellow}
\affiliation{Institute for Astronomy, 
    University of Hawaii,
    Honolulu, USA}
    
\author[0000-0001-7664-648X]{J. M. Joel Ong (王加冕)}
\altaffiliation{Hubble Fellow}
\affiliation{Institute for Astronomy, 
    University of Hawaii,
    Honolulu, USA}

\begin{abstract}
The stellar companions in the binary black hole systems Gaia BH2 and BH3, both of which are $\alpha$-enhanced red giant branch stars, are expected to show normal modes with the characteristic signature of convectively-driven solar-like oscillations. We investigate this using photometry from the TESS mission and find such a signal for Gaia BH2. For Gaia BH2, we measure a power excess frequency of $\nu_{\rm max}=60.15\pm0.57$ $\mu$Hz and a large separation of $\Delta\nu=5.99\pm0.03$ $\mu$Hz, yielding a mass of $1.19^{+0.08}_{-0.08}$ M$_\odot$, which is in agreement with spectroscopically derived parameters. Seismic modeling favors an age for the red giant of $5.03^{+2.58}_{-3.05}$ Gyr, strongly suggesting that it is a young, $\alpha$-enriched giant star, which are thought to arise from a binary accretion or merger scenario. Ground-based photometry of Gaia BH2 spanning 8 years indicates a photometric period of $398\pm5$ d, which we tentatively attribute to rotation. If this rotation is physical, it can not be explained solely by evolutionary spin-down or magnetic braking, and implies that the red giant underwent some tidal forcing mechanism. Suggestively, this period is close to the pseudo-synchronous spin period of P$_\text{spin}=428\pm1$ days derived from the binary orbit. For Gaia BH3, we are unable to identify an asteroseismic signal in the TESS data despite predicting that the amplitude of the signal should lie well above the measured noise level. We discuss a number of scenarios for why this signal may not be visible.

\end{abstract}


\section{Introduction}
\end{CJK*}

The recent astrometric discoveries of Gaia BH2 and BH3, the first binary systems consisting of a red giant star and a dormant black hole \citep{El-Badry2023Red, GaiaCollaboration2024Discovery}, have challenged our understanding of binary black hole formation \citep{Li2024Possible, Gilkis2024Gaia,Miller2024True,Tanikawa2024Compact,Kotko2024Enigmatic,Green2024Upper,MarinPina2024Dynamical}. In unveiling the formation mechanism of these systems, it is essential to accurately determine the fundamental properties (and in particular the mass and age) of the luminous component. This is because those of the compact component may often only be inferred from the luminous component, rather than being directly measurable. For example, if the orbits of these systems are characterized through astrometry, the reliability of the mass function (and therefore any estimate of the black-hole mass) crucially depends on the accuracy with which the stellar mass is measured.

Over the past century, variable stars have proven to be an invaluable resource in refining our understanding of stellar structure, evolution and fundamental stellar properties. Notably, the recent high-precision space photometry revolution, catalysed by the Kepler and TESS missions \citep{Ricker2015Transiting, Chaplin2015Asteroseismology, Guzik2016Detection}, has provided astronomers with an incredible amount of high-precision light curves spanning the entirety of the Hertzsprung-Russell diagram \citep[HRD;][]{Kurtz2022Asteroseismology}. The richness of these data has paved the way for asteroseismology -- the study of the internal structure of stars through their self-excited oscillations \citep{Aerts2010Asteroseismology,Aerts2021Probing}.

No class of star has benefited more from our improved understanding of asteroseismology than red giants \citep{Bedding2011Gravity}, in which stochastically excited modes have been identified with amplitudes around 0.1\,mmag and periodicities on the order of hours \citep[e.g.][for a review]{Hekker2018Asteroseismologyb, Basu2020Unveiling}. These oscillations are driven by near-surface turbulent convection (e.g., \citealt{Goldreich1977Solar, Balmforth1992Solar}), which both excites and damps modes around a characteristic power excess frequency ($\nu_{\rm max}$) \citep{tassoulAsymptoticApproximationsStellar1980}. Excited pressure modes (p-modes) of equal angular degree and different radial order are spaced in frequency by the large separation (\dnu) which is proportional to the mean density of the star \citep{Ulrich1986Determination}.  Such normal modes have been measured in tens of thousands of red giants \citep{Pinsonneault2014APOKASC, Pinsonneault2018Second,Yu2018Asteroseismology, Hon2021Quick, Mackereth2021Prospects}.

Asteroseismology has seen great success as a tool for characterizing other kinds of invisible stellar companions --- most notably exoplanets \citep[e.g.][]{Deal2017Asteroseismology, Lin2024Using, Huber2018Synergies, Lundkvist2018Using, Huber202220, Nielsen2020TESS}. We extend this now to the recently discovered dormant black hole systems by Gaia. In two of these systems, Gaia BH2 and Gaia BH3, the luminous companion has been conclusively shown to be a red-giant star \citep{El-Badry2023Red, GaiaCollaboration2024Discovery}. The precise mechanism by which these systems form remains a mystery. However, as effectively all sufficiently evolved red giants oscillate, we investigate these stars for their asteroseismic potential. The asteroseismology of these luminous companions may prove a promising source of clues about the under-specified physical nature of their formation and history.

In this paper, we produce light curves from the TESS target pixel files of the stellar companions to Gaia BH2 and Gaia BH3 (hereafter, BH2$^{*}$ and BH3$^{*}$). We find strong evidence of oscillations in BH2$^{*}$ at the frequency expected of stars with similar spectroscopic properties (Sec.~\ref{sec:bha_astero}), and a non-detection of oscillations in BH3$^{*}$ (Sec.~\ref{sec:bhb_astero}). For BH2$^{*}$, this enables a methodologically independent characterization of its fundamental properties, which we find to be in good agreement with existing published values. \rev{The age of \bha\ inferred from asteroseismology is younger than one would expect for an $\alpha$-enriched red giant in the solar neighborhood (i.e., $\sim 10$ Gyr; \citealt{2011MNRAS.414.2893F}). We also identify a photometric period in ground-based photometry which we tentatively suggest is caused by rotational modulation (Sec.~\ref{sec:bha_rot}). This period, the young asteroseismic age, and the fact that it is $\alpha$-enhanced suggest that it has undergone interactions with a companion, which we discuss in Sec.~\ref{sec:bha_merger}.} For BH3$^{*}$, a non-detection suggests that the uncertainties on the stellar parameters reported in the discovery paper are potentially too narrow, or that the scaling relations are inaccurate for very metal-poor stars (Sec.~\ref{sec:bhb_model}). In particular, for the published mass we expect an asteroseismic amplitude above the noise level of the light curve. We discuss several scenarios as to why the signal could not be found.

\section{Data and Methods}
\label{sec:method}

\subsection{Photometry}

We used data from the Transiting Exoplanet Survey Satellite (TESS; \citealt{Ricker2014Transiting}). TESS photometry is taken with pointings fixed for 27 day at a time, corresponding to one sector. The observing strategy of the nominal and first two extended TESS missions is such that only photometry in single sectors, separated by two years at a time, is available for both \bha and \bhb. For each sector of observation, we extract target pixel files using TESScut \citep{Brasseur2019Astrocut}.

We extract the light curves from the TESS target pixel files in each sector using a regression corrector as implemented in \texttt{lightkurve} \citep{viniciusKeplerGOLightkurve2018}. Briefly, the regression corrector operates by removing scattered light by detrending the light curve against vectors which are predictive of the systematic noise. These vectors are created from Principal Component Analysis (PCA) of pixels outside the selected aperture, and assumed to be representative of the scattered light. The regression corrected light curve is then normalized by dividing its median, and a 3$\sigma$ clipping is applied to remove significant outliers.

\subsection{Global Asteroseismic Properties and Scaling Relations}

The large sample sizes afforded by \kepler and \tess now allow the asteroseismic properties of stars, as a function of their stellar properties, to be understood to such an extent that we can predict the probability of detecting oscillations given a star's mass, temperature, surface gravity, while taking into account instrumental considerations such as cadence of observations and detector coordinates \citep{Schofield2019Asteroseismic}. 

Kepler observations have demonstrated that the oscillation amplitudes of red giants satisfy a scaling relation that depends on mass, luminosity and temperature \citep{Huber2011Testing, Corsaro2013Bayesian}:
\begin{equation}
    A_{\rm oscillation} \sim \frac{L^s}{M^t T_{\rm eff}^{r-1} c_K(T_{\rm eff})}
    \label{eq:ampl}
\end{equation}
where $s=0.838$, $t=1,32$, $r=2$, and 
\begin{equation}
    c_K \sim \left(\frac{T_{\rm eff}}{5934 K} \right)^{0.8}.
\end{equation}
Moreover, the large frequency separation $\dnu$ and frequency of maximum power excess $\numax$ of convectively-excited pulsations also scale with stellar properties like their masses, radii, and temperatures \citep{Brown1991Detection, Kjeldsen1995Amplitudes}:

\begin{equation}
\begin{aligned}
    \frac{\nu_{\rm max}}{\nu_{\rm max \odot}} &\sim \left(\frac{M}{M_\odot}\right) \left(\frac{R}{R_{\odot}}\right)^{-2} \frac{T_{\rm eff}}{T_{\rm eff \odot}}^{-1/2}\text{ and }\\
    \frac{\Delta\nu}{\Delta\nu_\odot} &\sim \left(\frac{M}{M_\odot}\right)^{1/2} \left(\frac{R}{R_\odot}\right)^{-3/2},
    \label{eq:scale}
\end{aligned}
\end{equation}
where the solar values are $\nu_{\rm max \odot} = 3090 \pm 30$ $\mu$Hz, $T_{\rm eff} = 5777$ K, and $\Delta\nu_\odot = 135.1 \pm 0.1$ $\mu$Hz.

Our observational characterization of these global seismic properties described below will rely primarily on existing automated software products, such as the \pysyd pipeline \cite{Chontos2021PySYD} and \pbjam \citep{nielsen_pbjam_2021}.

\section{Gaia BH2}
\label{sec:bha}

\begin{figure*}[t]
    \includegraphics[]{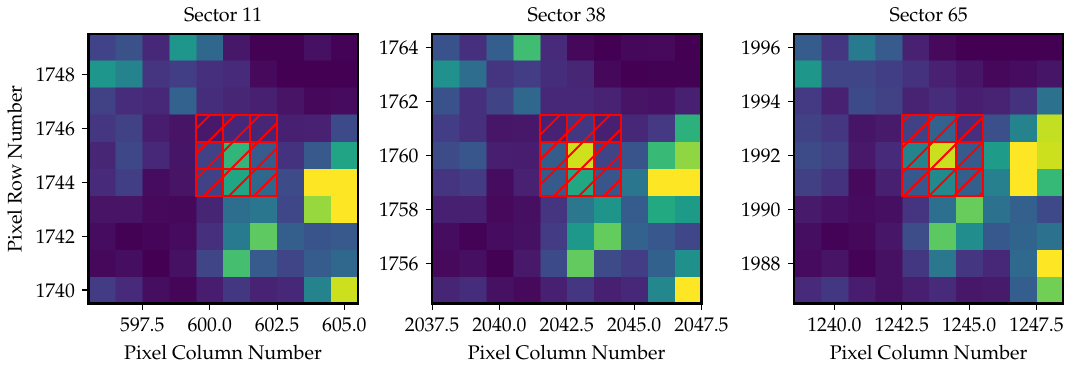}
    \includegraphics[width=\linewidth]{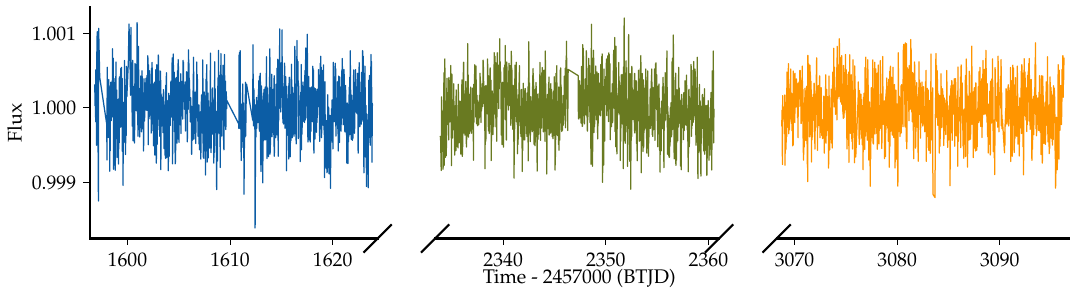}
    \caption{Top panel: TESS Target pixel file cutouts of the three sectors of Gaia BH2, with the selected aperture overlaid in red. Note the presence of potential nearby contaminants. Within two pixels of the selected aperture, there are no contaminants brighter than 13th Gaia magnitude beside \bha itself. Bottom panel: Light curves from each sector obtained through regression correction. \rev{All the light curves have been binned to a uniform 30-minute cadence, and outliers have been removed.}}
    \label{fig:bh2_reduction}
\end{figure*}

\begin{figure*}
    \centering
    \includegraphics[]{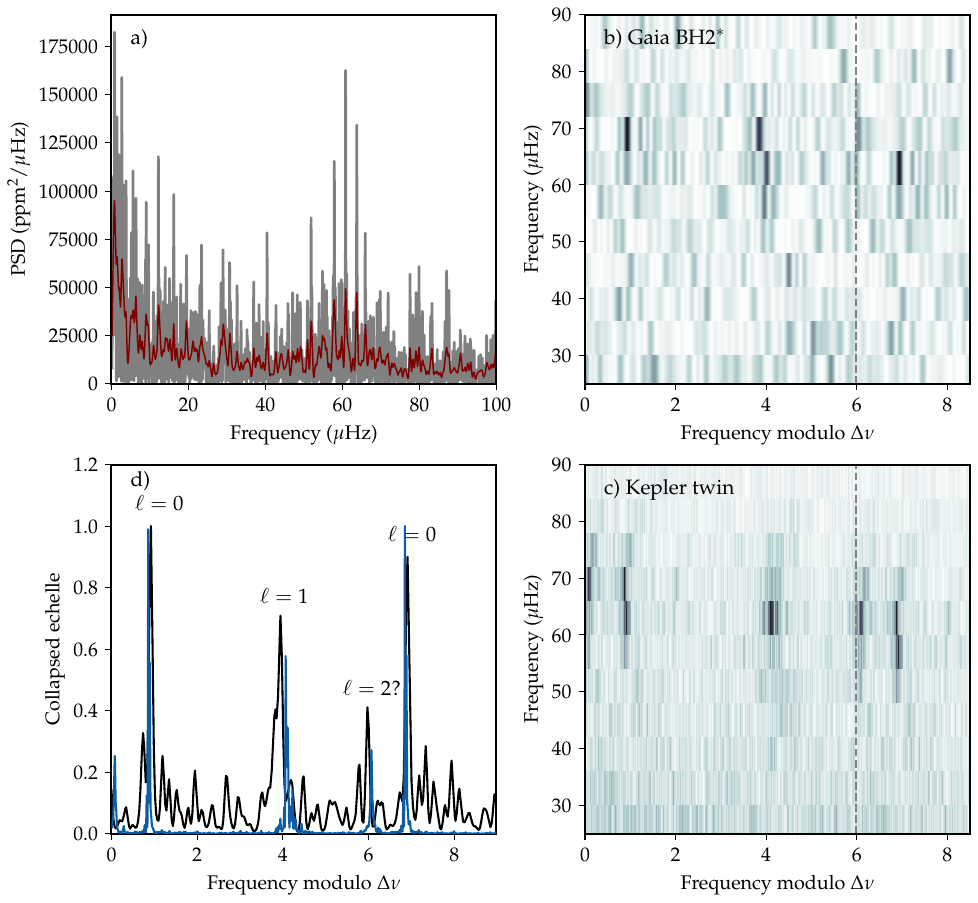}
    \caption{\textbf{a)}: Power spectral density of \bha, with the power excess around 60$\mu$Hz. The red line is the power spectrum convolved with a Gaussian kernel of width 0.2 $\mu$Hz, highlighting the individual modes. \textbf{b)}: The echelle diagram of the PSD at the measured \dnu\  (dashed line). \textbf{c)}: The echelle diagram for a Kepler twin (see text), with similar stellar properties. \textbf{d)}: \rev{The collapsed echelle diagram of \bha (black) and the Kepler twin (blue), highlighting the radial, dipole, and quadrupole modes.\label{fig:bh2psd}}}
\end{figure*}

\bha (Gaia DR3 5870569352746779008, TIC 207885877) is a G=12.3 magnitude nearby (d=1.16 kpc) star on the lower red giant branch, before the red clump. \citet{El-Badry2023Red} infer a mass of M$_* = 1.07 \pm 0.12$ M$_\odot$, moderately subsolar metallicity ([Fe/H] = -0.22$\pm$0.02) and strong $\alpha$ enhancement ([$\alpha$/Fe]=0.26$\pm$0.05), by fitting spectroscopic observations against evolutionary models. They suggest that the strong $\alpha$-enhancement is primordial owing to the wide orbit of the system, but could also arise from pollution of low-velocity ejecta during a failed supernova. 

\bha was observed by TESS in sectors 11, 38, and 65 at cadences of 30 minutes, 10 minute, and 2 minutes respectively. Fig.~\ref{fig:bh2_reduction}. We use a 10x10 TESS pixel cutout, and show the target pixel file and light curve in Fig.~\ref{fig:bh2_reduction}. The region surrounding \bha is moderately crowded, meaning that we must be careful when choosing an aperture. \rev{We experiment with multiple apertures centred on \bha, permuting through every combination up to a 3x3 aperture. This permutation is performed on every possible pixel combination that includes the central and adjacent pixels. We measured SNR of the power excess for each permutation, calculated as the ratio of the dominant mode amplitude against the amplitude at high frequencies.} We find that the 3x3 aperture best captures the asteroseismic signal with the highest SNR. Importantly, the signal shows up in every aperture which is centered around \bha, and is unlikely to belong to a nearby target. The closest target of significance to \bha is Gaia DR3 5870569318392555776, an A/F-type star which would be expected to have significantly higher amplitude and frequency of variability. \rev{We join the light curves together by normalizing each sector, binning to 30-minute cadence, and stitching them together. Note that sector 11 is already observed in 30-minute cadence, with sectors 38 and 65 observed at 10-minute cadence. We separately investigate the higher frequency regions probed by the 10-minute cadence light curves before binning, and find no additional signals of interest.}

\subsection{Asteroseismology}
\label{sec:bha_astero}

\begin{table}[]

\begin{tabular}{lcr@{}}
\multicolumn{3}{l}{\textit{Gaia BH2}} \\
\toprule
\multicolumn{3}{l}{\textbf{Measured quantities}} \\
Power excess                 & $\nu_{\rm max}$ ($\mu$Hz)   & $60.15\pm0.57$ \\
Large separation             & $\Delta\nu$  ($\mu$Hz) & $5.99\pm0.03$ \\
Small separation             & $\delta\nu_{\rm 0,2}$  ($\mu$Hz) & $0.93\pm0.05$ \\
Photometric period & P$_{\rm phot}$ (day) & $398 \pm 5$ \\
                   &                   &  \\
\multicolumn{3}{l}{\textbf{Literature quantities}$^\dagger$} \\
Temperature                  &   T$_{\rm eff}$ (K)                & $4604\pm87$ \\
Bolometric luminosity & L$_*$ (L$_\odot$) & $24.6\pm1.6$\\
Surface gravity & $\log{g}$ (cm s$^{-2}$)& $2.71\pm0.24$\\
Radius &R$_*$ (R$_\odot$)  & $7.77\pm0.25$\\
Mass & M$_*$ (M$_\odot$)  & $1.07\pm0.19$\\
Metallicity &[Fe/H] &  $-0.22\pm0.02$\\
$\alpha$-enhancement & [$\alpha$/Fe]& $0.26\pm0.05$\\

                   &                   &  \\
\multicolumn{3}{l}{\textbf{Derived quantities (\texttt{asfgrid})}} \\
Radius                   &    R$_*$ (R$_{\odot})$               &  $8.55\,^{+0.20}_{-0.15}$\\
Mass                         & M$_*$ (M$_{\odot})$                 & $1.23\,^{+0.09}_{-0.09}$ \\
Age &$\tau$ (Gyr) & $5.11\,^{+1.22}_{-1.78}$ \\

\multicolumn{3}{l}{\textbf{Derived quantities (\texttt{cnfgiant})}} \\
Mass                         &  M$_*$ (M$_{\odot})$                 & $1.19\,^{+0.08}_{-0.08}$ \\
Age &$\tau$ (Gyr) & $5.03\,^{+2.58}_{-3.05}$ \\

\bottomrule
\end{tabular}
\caption{Measured, derived, and literature values of the red giant in the binary system Gaia BH2. Note that the radius is fixed for the \texttt{cnfgiant} grid because the luminosity and temperature are inputs. $^\dagger$From \citet{El-Badry2023Red}. \label{tab:bh2}}
\end{table}





The power spectrum of \bha is shown in Fig.~\ref{fig:bh2psd}, where we find a power excess corresponding to solar-like oscillations at frequencies between 50 to 80 $\mu$Hz. The photometric data has a white noise level of 22 ppm at high frequencies. We use the \pysyd pipeline to measure a \numax\ of $60.15\pm0.57$ $\mu$Hz and a large frequency separation \dnu\ of $5.99\pm0.03$ $\mu$Hz. \rev{Briefly, \pysyd follows the same routines as the original \textsc{SYD} pipeline (cf. \citet{Huber2009Automated}. Although largely automated, we set the upper and lower bounds when calling \pysyd to be in the range from 20 to 100 $\mu$Hz. We also use the MCMC option to estimate the uncertainties on \numax\ and \dnu.}

We also attempted to use \pbjam for this purpose, which should in principle both be capable of deriving more precise estimates of \dnu\xspace and \numax\ by fitting a template power spectrum of even-degree p-modes, as well as of estimating the frequencies of individual normal modes. This procedure yielded $\dnu=6.01 \pm 0.02\ \mu\mathrm{Hz}$ and $\numax=61 \pm 1\ \mu\mathrm{Hz}$. However, the adverse S/N conditions of single-sector TESS data prevented PBJam's automated procedures from deriving a robust mode identification, since it relies on the visibility of quadrupole modes to distinguish between the most prominent $\ell = 0$ and $\ell = 1$ modes. Moreover, since this template-fitting procedure is constrained only by even-degree modes, it has less statistical support than would be obtained from also including the dipole modes. As such, we adopt the values reported by \pysyd for our subsequent analysis.

Rather than resorting to mode identification by eye for such a low S/N power spectrum, we turn to lessons about the behavior of pulsating red giants that the asteroseismology community has learned from the Kepler sample. There are over 16,000 red giant stars observed by Kepler with measured asteroseismic properties \citep[e.g.][]{Yu2018Asteroseismology}. The majority of these stars have also been observed by TESS in multiple sectors and cadences. Using this, we can identify ``twins" in the Kepler sample for any given red giant: Kepler targets with a similar apparent magnitude and $\nu_{\rm max}$ as predicted from the published stellar parameters. We compared \bha against all Kepler red giants in the consolidated APOKASC3 catalogue \citep{Pinsonneault2024APOKASC3}, quantifying similarity using the $\chi^2$ statistic, incorporating constraints from metallicity, effective temperature, and $\Delta\nu$. This analysis identifies KIC 2712761 as the star with the lowest $\chi^2$ discrepancy from \bha. We will refer to it as \bha's ``Kepler twin''.

The frequency echelle diagram of the power spectrum of \bha in comparison to its Kepler twin (Fig.~\ref{fig:bh2psd}) shows the oscillation modes forming three distinct ridges we consider to be the $\ell=0$, $\ell=1$, and $\ell=2$ modes. This is corroborated by relying on relations between \dnu\ and the p-mode phase offset $\epsilon$ \citep{White2011CALCULATING,Ong2019Structural}. We measure the separation between the $\ell=0$ and $\ell=2$ ridges (the small separation; $\delta\nu_{0, 2}$) using the collapsed echelle with a Monte-Carlo routine, finding a value of $0.93 \pm 0.05$ $\mu$Hz. 

\citet{El-Badry2023Red} similarly extracted TESS light curves for \bha in search of asteroseismic signatures, finding a weak detection of a signal with a peak frequency at 61 $\mu$Hz. While we confirm these results here, we are also able to further identify the large and small separations. The difference between their result and our unambiguous detection of oscillations is most likely due both to selection of the aperture and light curve extraction, as well as our having additional data at our disposal: at the time of their analysis, \citet{El-Badry2023Red} did not have access to Sector 65 observations taken at 10 minute (rather than 30 minute) cadence, which greatly improves the S/N of the power excess.

To obtain stellar properties from our measured \numax\ and \dnu, we use the normalizing flow prescription outlined in \citet{Hon2024Flowbased} using the software \textsc{modelflows}, based on \texttt{asfgrid} stellar models with corrected large frequency separations for lower metallicity \citep{Sharma2016LAR, Stello2022Extension}. In this method, stellar models are emulated using normalizing flows, allowing us to sample the posterior mass and radius of the star with our input measurements. Given that \bha is $\alpha$-enhanced ([$\alpha$/Fe]=$0.26\pm0.05$), we correct the input metallicity using the prescription of \citet{Salaris1993Alpha} and obtain a corrected metallicity of $=-0.037\pm0.04$ dex. 

Using our values from Table.~\ref{tab:bh2} and the measured temperature, we sample the stellar grid (\texttt{asfgrid}) for 5000 points with `evstate' set to zero, corresponding to the red giant branch. We also make use of the \texttt{cnfgiant} grid available in \textsc{modelflows}, which is a narrower grid defined only for hydrogen shell-burning stars with $0.7 \leq M \leq 2.5 M_\odot$. This grid allows us to incorporate the small separation and luminosity as constraints, which effectively fixes the radius of the star. We note that the small separation is significantly weaker as a diagnostic of internal structure for stars on the red giant branch compared to the main sequence \citep{Ong2025Resolving}, and so has little effect on our results -- running the same grid without the small separation results in effectively identical results.

For the \texttt{asfgrid} models we find a mass of $1.23^{+0.09}_{-0.09}$~M$_\odot$ and radius of $8.55^{+0.20}_{-0.15}$~R$_\odot$. For the \texttt{cnfgiant} models we find a mass of $1.19^{+0.08}_{-0.08}$~M$_\odot$. Both these masses are within 1$\sigma$ of the literature value \citep{El-Badry2023Red}, but the radius is slightly larger. Table~\ref{tab:bh2} summarizes these results.

The larger uncertainty in age for the \texttt{cnfgiant} grid is due to the larger number of free parameters, such as initial helium abundance, mixing length parameter, and convective core and envelope overshooting, which are unconstrained in this instance. The difference in ages between model grids is only 0.08 Gyr, and so we consider the systematic uncertainty to be very small. We adopt the age from the \texttt{cnfgiant} grid, which is produced solely for red giants; $\tau=5.03^{+2.58}_{-3.05}$ Gyr. The model age confirms that \bha is a young, $\alpha$-enhanced star, if we assume no departures from canonical single-star evolution. We discuss these implications in Sec.~\ref{sec:bha_merger}.

\subsection{Long-term photometric modulation}
\label{sec:bha_rot}

\begin{figure*}
\centering
    \includegraphics[]{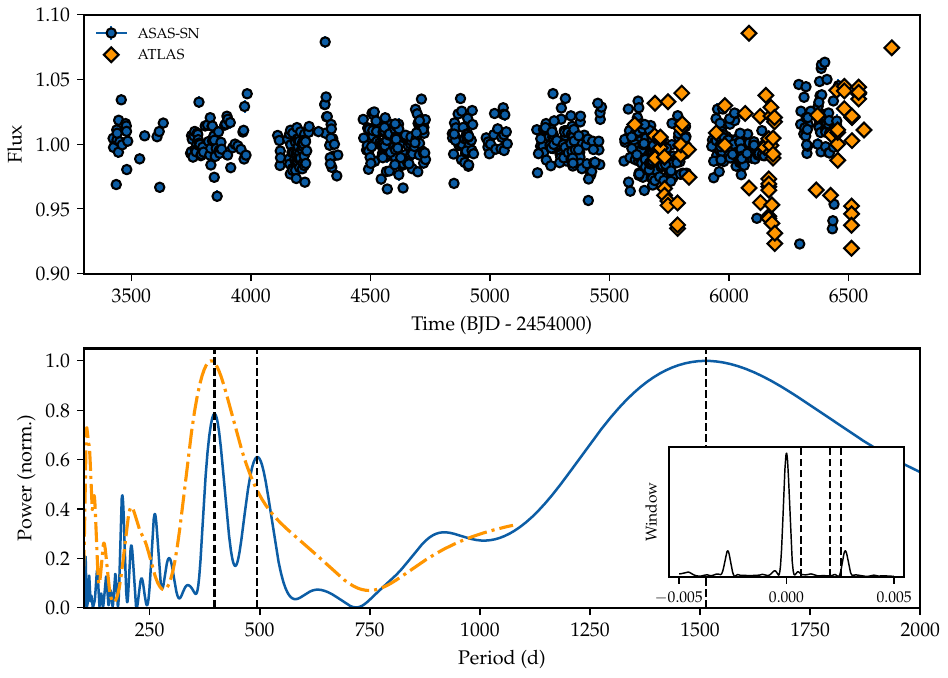}
    \caption{\rev{Top panel: ASAS-SN (blue circle) and ATLAS (orange diamond) light curves of \bha. The ASAS-SN light curve is the combination of several different cameras in the network, each of which have their own offset which has been median divided out. Bottom panel: The power spectra of the ASAS-SN (blue) and ATLAS (orange) light curves. The dashed black lines indicate measured periods at 397 d, 495 d, and 1513 d. The inset on the bottom panel shows the window function (in frequency, d$^{-1}$), with a signal at Earth's orbital frequency (1/365) d$^{-1}$. There are no peaks in the ASAS-SN window function associated with our measured periods (black dashed lines).\label{fig:bh2asassn}}}
\end{figure*}

\begin{figure}
    \centering
    \includegraphics[width=.495\textwidth]{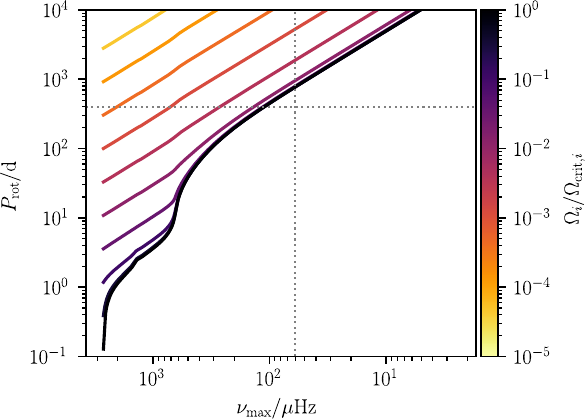}
    \caption{The photometric rotational period of \bha is unlikely to have arisen without tidal forcing. Here we show the rotational evolution of \textsc{mesa} stellar models initialized at various fractions of the ZAMS breakup rotational frequency, indicated by the colourmap. Magnetic braking \citep[calculated here using the prescription of][]{vansaders_rotation_2013} causes stellar rotation to slow over evolutionary timescales, shown here using $\numax$ as an age proxy. No other sources of angular momentum loss (such as mass loss) are accounted for in this simulation. Even under this extremely permissive scenario, our putative photometric rotation period (horizontal line) at the present day (vertical line) is faster than permitted by if \bha's rotation were to be attributed entirely to its primordial angular momentum. \label{fig:braking}}
\end{figure}

While relatively uncommon, many red giant stars are known to show long-term photometric modulation associated with rotation (e.g., \citealt{Ceillier2017Surface, Gaulme2014Surface, Gaulme2020Active, Ong2024Gasing}). Given that TESS's short-sector observing strategy and instrumental systematics render it difficult to constrain signals with periods longer than the length of contiguous sectors \citep{Colman2024Methods, Claytor2024TESS, Binks2024TESSILATOR}, we instead determine these using the ASAS-SN (the All-Sky Automated Survey for Supernovae; \citealt{Kochanek2017AllSky}) light curves of \bha. ASAS-SN is a ground-based network of observatories dedicated to transient science and long-term photometric variability. \bha is in the ideal magnitude range for observations with ASAS-SN \citep{Hart2023ASASSN}.

To separate true signals from potential systematics associated with ASAS-SN's observing strategy and instrumentation, we also produce a light curve using the Asteroid Terrestrial-impact Last Alert System \citep[ATLAS:][]{tonry_atlas_2018}. ATLAS is a ground-based mission with effectively daily cadence across the entire sky, with a pixel scale of 1.86". Although primarily a mission focusing on the northern hemisphere, ATLAS now also has cameras stationed in the southern hemisphere which can image stars at lower declination. We produce both difference and reduced photometry of \bha using the ATLAS forced photometry server\footnote{\url{https://fallingstar-data.com/forcedphot/}}, which provides public access across the entirety of the ATLAS mission. Data processing and photometry are described in more detail in \citet{Tonry2018ATLAS} and \cite{Smith2020Designa}. We also obtain approximately 8 years of ASAS photometry (ASAS-3; \citealt{Pojmanski1997All}) in the V-band. Manual inspection of the ASAS-3 data suggests a period of 426 days, however, the false alarm level of the signal is over 10\% and so we do not include it in our analysis.

The ASAS-SN light curve of \bha has an average uncertainty on each point of 0.005 mag. The light curve shows three clear periodicities at approximately 1500 d, 390 d, and 494 days above a 1\% false alarm level (listed in order of decreasing amplitude) (Fig.~\ref{fig:bh2asassn}). Given that the survey is ground-based, it is important to ensure that this signal is not a systematic artifact of the survey, such as being an alias of the Earth's diurnal, annual, or lunation frequencies. We therefore inspect the observational window function for the ASAS-SN power spectrum, shown inset in the bottom panel of \autoref{fig:bh2asassn}. We find that these three periodicities do not correspond to any sidelobes of this window function, suggesting that they are astrophysical rather than systematic in origin. We also perform the same analysis after applying Gold deconvolution to suppress sidelobes in the spectral window function of the data \citep{Morhac2006Deconvolution, Li2025Kdwarf}. The deconvolved spectrum shows the same periodicities. We measure these periods by fitting a sine wave with a flux offset to the light curve using \textsc{emcee} for 5000 tuning steps and 5000 draw steps. We fit only the ASAS-SN light curve, because the ATLAS data has significantly larger scatter. We find periods of $398 \pm 5$, $494 \pm 9$, and $1513 \pm 65$ days.

Given the crowded field, it is again possible that the signal is contaminated from nearby stars even though ASAS-SN has a smaller pixel scale (8") than TESS (21"), and ATLAS is even smaller (1.86"). There are 19 stars within 20" of \bha according to Gaia DR3, with the brightest companion at 16th Gaia magnitude, in comparison to \bha itself which is at \rev{approximately 12th Gaia magnitude}. The remaining majority of companions are around 20th Gaia magnitude, which would be too faint to measure photometric variations with ASAS-SN. We construct light curves of nearby stars and find no matching signal at the photometric periods we measure.

In summary, both the ASAS-SN and ATLAS data products indicate the presence of quasi-periodic photometric variability with a common period of $P_\text{phot} = $ \bharot. This is far slower than both the fundamental period of \bha's oscillations ($\sim 0.2$ d), as well as the correlation timescale of convective granulation in its envelope ($\sim 1$ d; \citealt{Kallinger2014Connection}). Rather than stellar-astrophysical processes, we instead investigate if this might be attributable to any of the characteristic timescales associated with \bha's tidal interactions with its black hole companion. For example, the presence of a binary companion in a close orbit might endow it with faster rotation than we might ordinarily expect from a comparable isolated red giant \citep{zahn_tidal_1977}. 

Notably, \bha is in an eccentric orbit with $e = 0.5176 \pm 0.0009$. Under such circumstances, \cite{hut_tidal_1981} describes how tidal torques might spin \bha up to rotate more quickly than the orbital period: these torques would be much stronger near periastron (where the orbital frequency is fastest) than at apoastron. Thus, the average torque over the course of the orbit arises primarily from contributions close to periastron. As such, in addition to the mean orbital period ($P_\text{orb} = 1276.7 \pm 0.6 \ \mathrm{d}$), this yields two other characteristic periods:
\begin{itemize}
    \item The orbit-averaged torque vanishes when \bha rotates at the ``pseudosynchronous'' period
    \begin{equation}
        P_\text{pseud} = {\left(1 + 3e^2 + {3 \over 8} e^4\right)\left(1 - e^2\right)^{3\over 2} \over 1 + {15 \over 2} e^2 + {45 \over 8} e^4 + {5 \over 16} e^6} \cdot P_\text{orb},
    \end{equation}
    rather than at the orbital period. Evaluating this expression with the reported values of the orbital period and eccentricity yield $P_\text{pseud} = 428 \pm 1 \ \mathrm d$.
    \item Without assuming that \bha's rotation has been captured into pseudosynchronous resonance, we also note that its rotation period is strictly bounded from below by the rotational period associated with the angular frequency at periastron,
    \begin{equation}
       P_\text{min} = {(1-e^2)^{3\over2}\over(1+e)^2} P_\text{orb} = 347 \pm 1\ \mathrm{d}.
    \end{equation}
\end{itemize}
We find that $P_\text{phot}$ is of intermediate value between the minimum and pseudosynchronous period, and much closer to the latter than the former: this is highly suggestive of an interpretation of it being \bha's rotation rate. Such a fast rotation rate would be compatible with \bha being either a single star or a merger remnant. In the single-star scenario, this rotation must necessarily result from tidal forcing, since it is too fast to have originated from its primordial angular momentum (\autoref{fig:braking}). Conversely, if \bha were to have been a merger remnant, this would result from magnetic braking stalling when the rotation rate comes into pseudosynchronisation.

In either case, if indeed this quasi-periodic variability should be attributable to stellar rotation, this would also explain the relatively low visible amplitudes of the non-radial modes that we observe in \bha, compared to the radial modes. Photometric variability would only trace rotation through the emergence of features on the stellar surface, such as spots originating from magnetic activity, which produce photometric modulations as they are rotated onto and off the visible disc. Such magnetic fields would have the effect of stabilizing convective flows, as does the rotational motion itself; both effects are known to suppress the oscillation amplitudes \citep{Bonanno2014Asteroseismic,Bonanno2019Acoustic,Gaulme2020Active,Corsaro2024New}, and more strongly at higher angular degree.

The 10-year ASAS-SN data set also exhibits a second peak in its power spectrum, at $P_\text{long} \sim 1500\ \mathrm d$, which may also be a candidate rotation period. However, it is less easy to explain how such a rotational period could have arisen, since it is longer than the present orbital period, and does not correspond to either of the equilibrium timescales described above. To see why, we note that in addition to these, we in principle also have a hierarchy of secular-evolution timescales, associated with rotational (pseudo)synchronisation, tidal inspiral, and tidal circularization, ordered as $t_\text{sync} \sim (a / R)^6 < t_\text{decay} \sim (a/R)^{13/2} < t_\text{circ} \sim (a / R)^8$ \citep{zahn_tidal_1977,hurley_evolution_2002,lai_tidal_2012}.

Achieving a rotation period of $P_\text{long}$ could be possible under the single-star scenario if we were to demand that $t_\text{sync}$ were marginally faster than the rotational braking timescale of roughly $10^7\ \mathrm{yr}$, in order to spin it up tidally beyond the otherwise negligible rotation rates that it would ordinarily possess this far up the red giant branch, while also avoiding capturing it into pseudosynchronisation. However, since $t_\text{decay} \sim 10 t_\text{sync}$, this would also imply in-spiral on timescales comparable to that of evolution up the red giant branch, of roughly $10^8\ \mathrm{yr}$. Coupling \bha to tidal spin-up as a canonical single star, while avoiding runaway inspiral, would therefore only be possible with considerable fine-tuning of these timescales. If instead \bha were to be a merger remnant, we would instead require $t_\text{sync}$ to be much longer than the braking timescale, so that the initially rapid rotation after the merger does not stall in its braking when near pseudosynchronisation. However, since mass transfer rapidly circularizes orbits, this would require there to historically have been essentially no mass transfer from the black hole progenitor to \bha, or its own progenitors.

Either way, while one would identify the additional appearance of $P_\text{phot}$ as being the 4th harmonic of the rotational period if $P_\text{long}$ were to signify rotation, it would be difficult to explain why other lower harmonics are not present in the power spectrum. Conversely, if we were to identify $P_\text{phot}$ as the rotation period, a longer-period signal might be attributable to the timescales over which the spot morphologies change, or to latitudinal differential rotation.

As such, we believe it would be more plausible to interpret $P_\text{phot}$, rather than $P_\text{long}$, as being representative of \bha's present rotation rate. If this period is indeed the rotational period of the star then it corresponds to an equatorial velocity of $v_{\text eq}=0.99$ kms$^{-1}$ and equivalent $v\sin{i}$ assuming the star is seen equator-on. The influence in the line-broadening of the spectrum of the star would be minimal to non-existent, especially considering that \citet{El-Badry2023Red} model an upper limit on $v\sin{i}$ of $<1.5$ kms$^{-1}$. Otherwise, we defer a more detailed investigation of possible secular evolution to future work when more photometric data is available.

\subsection{Is Gaia BH2$^*$ a merger product?}
\label{sec:bha_merger}

Our asteroseismic model age requires that the red giant in Gaia BH2 belongs to the class of young ($<8$ Gyr), $\alpha$-enriched red giants in the solar neighborhood. Such stars show an unusually high abundance of $\alpha$ elements, characteristic of thick disk stars, but are otherwise known to be young \citep{Martig2015Young, Chiappini2015Young, SilvaAguirre2018Confirming, Claytor2020Chemical, Das2020Ages,Zinn2022K2, Pinsonneault2024APOKASC3, Zhang2021Most}. One possible interpretation of these stars is that they may be products of mass transfer or merger events. That is, they have a higher mass than they began with, and so appear younger when their ages are estimated assuming canonical single-star evolution \citep{Zhang2021Most,Yu2024New, Lu2024Evidence, Grisoni2024K2}. \citet{Miglio2021Age} suggest that the occurrence rate of massive ($>1.1 M$ sun) alpha rich red giants is on the order of 5\% for the RGB, and significantly higher in the RC. \rev{We discuss now some possible origins of these $\alpha$ elements.}

\rev{\citet{El-Badry2023Red} consider the possibility of \bha having been chemically enriched through pollution from the supernova of the BH progenitor, since such enhancement is also seen in the donor stars of BH X-ray binaries. They argue, from geometric considerations, that the observed $\alpha$-enhancement is unlikely to have arisen by the accretion of supernova ejecta. Instead, they suggest it is possible that low velocity ejecta (such as in Type IIn supernovae, where ejecta interacts with dense circumstellar material; \citealt{Taddia2013Carnegie})) remained gravitationally bound to the star and eventually accreted onto \bha.  }

\rev{It is also possible that \bha was enriched by interactions with the BH progenitor star itself, prior to core-collapse. For an assumed progenitor mass of $>25$~M$_\odot$ \citep{Sukhbold2016CORECOLLAPSE, Raithel2018Confronting},} such a star would exceed the current orbital separation of the binary ($4.96$ AU). This common envelope evolution is disfavored, as it significantly reduces the orbital separation to much lower than what it is today \citep{El-Badry2023Sunlike}.

Merger remnants, or planetary engulfment, are also one explanation for the existence of rapidly-rotating red giants with no obvious stellar companions \citep[e.g.][]{phillips_seven_2023,Ong2024Gasing,rui_finding_2024}. Tidal synchronization (which we discuss above) may serve to brake initially rapid rotation more quickly than would magnetism alone, while at the same time also producing extra tidal heating that may continue to puff the star up even after after many Kelvin-Helmholtz timescales have elapsed.

\rev{This being so, it is tempting --- considering the asteroseismic age of \bhaage, its $\alpha$-abundance, and apparently rapid rotation --- to suggest that \bha has undergone significant interactions with a past companion: either the progenitor to the present black hole, or another star. Briefly, we simulate the difference in mass for a young (5 Gyr) and old (8 Gyr) $\alpha$ enhanced RG with the same properties as \bha\ using stellar models drawn from \textsc{asfgrid}. We find that to make \bha\ appear young requires around 0.25 M$_\odot$ of additional mass, far above the limit of a substellar companion.} 

In addition to these rotational or kinematic tracers of mergers or accretion, asteroseismic signatures of them also exist \citep[e.g.][]{li_discovery_2022,Henneco2024Merger}: further asteroseismic measurements, with longer data sets, may prove helpful in determining whether \bha has undergone such a previous history. For example, while we have been able to detect standing acoustic waves (p-modes) with the TESS data that we have in hand, all nonradial p-modes are known to couple to standing buoyancy waves (g-modes) that are otherwise trapped in the radiative cores of red giants \citep[e.g][]{shibahashi_modal_1979,unno_nonradial_1989,Beck2011Kepler}, to produce gravitoacoustic mixed modes. 

We have so far only been capable of measuring the most p-dominated of these mixed modes in \bha. \kepler data show that, in addition to these p-mode features, a sufficiently long observing baseline might enable the measurement of the dipole g-mode period spacing $\Delta\Pi_1$ in comparably evolved stars \citep[e.g.][]{vrard_period_2016}, among other features. Remarkably, single stars are bounded by a tight sequence in the $\Delta\nu$-$\Delta\Pi_1$ diagram that is set by core electron degeneracy \citep{deheuvels_seismic_2022} --- we might therefore expect \bha to lie on this sequence, given that its mass suggests it possesses an electron-degenerate core. If a red giant should lie in the forbidden region on this diagram --- i.e. be more puffed up than expected of an ordinary red giant with a given electron-degenerate core size --- then this is thought to be one indicator that its exterior may have been modified by physical processes not ordinarily accounted for in canonical stellar evolution, as would indeed be the case in the event of a merger \citep{Rui2021Asteroseismic} or mass accretion. As such, a period-spacing measurement would afford us a model-independent diagnostic of such noncanonical evolution, and serve as a test of whether we ought to consider this model age to be reliable. More photometric data could thus conclusively determine the evolutionary history of \bha. 

\section{Gaia BH3}


\begin{figure*}
    \centering
    \includegraphics[]{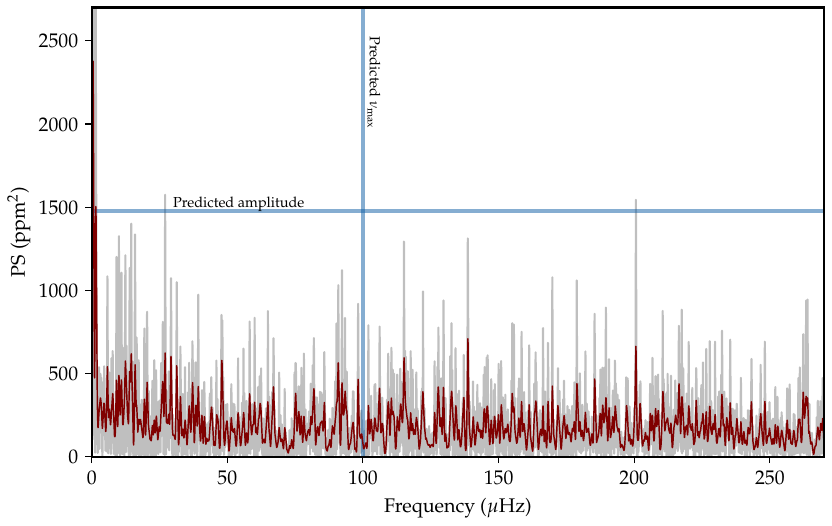}
    \caption{Power spectrum of \bhb. The red line is the power spectrum convolved with a Gaussian kernel of width 0.2$\mu$Hz, to highlight the absence of any significant signal. The blue vertical and horizontal lines correspond to the predicted power excess and power of the asteroseismic signal. \label{fig:bh3psd}}
\end{figure*}

\bhb (Gaia DR3 4318465066420528000, TIC 73788466) is a high proper motion star, with its absolute magnitude and color identifying it as ascending the red giant branch. In  \citet{GaiaCollaboration2024Discovery}, it was discovered to be host to a dormant 33$M_\odot$ black hole in a wide ($P=4195\pm112\ \mathrm d$) eccentric (e=0.73) orbit, with the black hole having extremely low near-infrared emission \citep{Kervella2025VLTI}. \bhb has been observed by TESS in two non-contiguous sectors, 54 and 81 at a cadence of 10 and 2 minutes respectively.

The stellar abundance analysis of \bhb identified it as extremely metal-poor ($\feh = -2.56\pm0.11$) and moderately $\alpha$-enhanced ($\afe = 0.43\pm0.12$). Interestingly, the authors note a lack of $_{13}C$ and solar levels of [Ba/Fe], indicating that the star has not been chemically enriched by a companion star in the AGB phase. Given that it has evolved in apparent isolation, the asteroseismic properties of the star should be normal, conforming to what we understand of how these properties vary over time.

\rev{The star itself was modeled by \citet{GaiaCollaboration2024Discovery}.} Notably, the reported stellar mass ($0.76\pm0.05 M_\odot$) is extremely precise given that it was derived through isochrones. Since the stellar evolution models of stars clump up on the red-giant branch, and as noted by \citet{Tayar2022Guide}, isochrone uncertainties in mass for red-giant stars are typically much larger. We show, later on, that this mass uncertainty is almost entirely due to the tight age and luminosity priors placed on the star.

Given the moderately crowded region and high proper motion of Gaia BH3, we extract a 20x20 target pixel file for both sectors of observation. We follow the same regression correction method outlined in Sec.~\ref{sec:method} with a 3x3 aperture, PCA comprising six components, and bin the light curve to a 30 minute cadence.

\begin{figure*}
    \centering
    \includegraphics{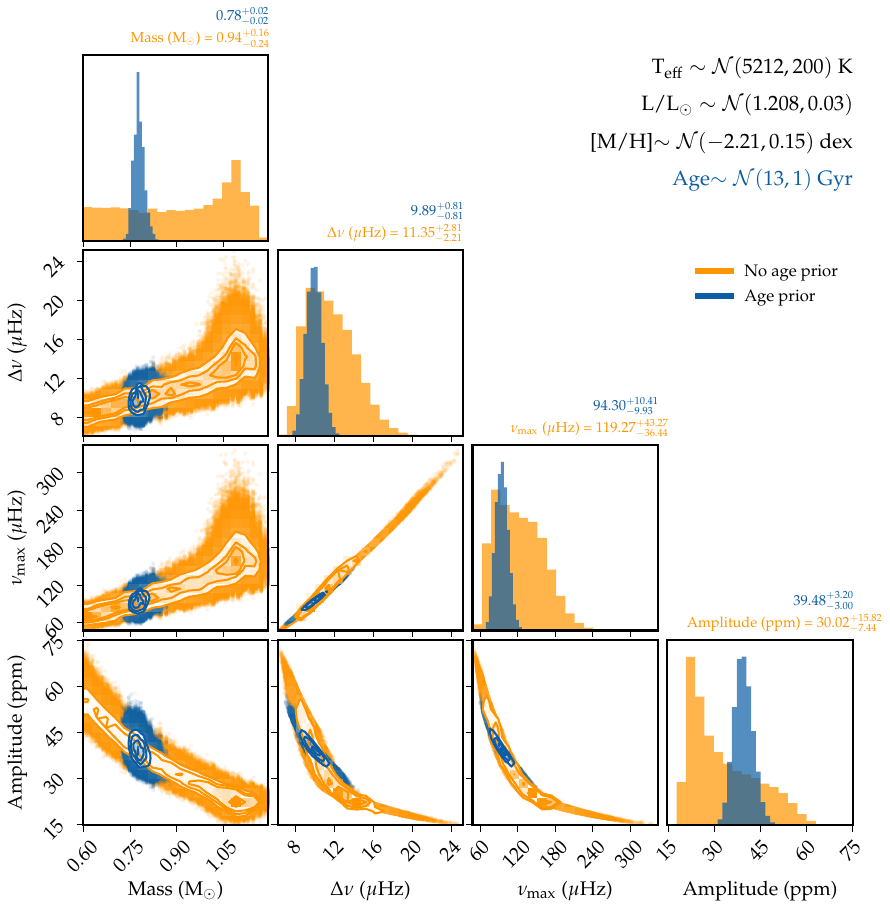}
    \caption{Posteriors of stellar properties of \bhb sampled from the stellar models, with temperature, luminosity and metallicity priors (orange), and with an additional age prior (blue).}
    \label{fig:model}
\end{figure*}

\subsection{Asteroseismology}
\label{sec:bhb_astero}

\bhb shows no obvious signs of asteroseismic variability, even after permuting both the aperture and number of PCA components (Fig.~\ref{fig:bh3psd}). There is a small increase in noise around 100\,$\mu$Hz which could possibly be interpreted as stochastic variability. We discuss now the absence of this signal, and possible causes.

As in \autoref{sec:bha}, we search for Kepler light curve twins for \bhb. We compare it against the twin based on the implied \numax\ and \dnu\ obtained from the asteroseismic scaling relations (Eq.~\ref{eq:scale}). The goal of this exercise it to demonstrate that for the measured stellar parameters of \bhb in \citet{GaiaCollaboration2024Discovery}, the TESS light curve of a known Kepler star at a similar magnitude should show a clear signal. 

Using the stellar parameters in \citet{GaiaCollaboration2018Gaia}, under the assumption that they are normally distributed, we predict $\nu_{\rm max}=100.3\pm1.0 \mu {\rm Hz}$. However, in metal-poor stars the \numax{} scaling relation can be systematically off by up to 20\% (e.g., \citealt{Huber2024Stellar,Chaplin2020Age,LiTanda2022}). In \citet{Yu2018Asteroseismology}, there are 83 Kepler stars within 1$\sigma$ of this value at various magnitudes, and one Kepler star at a similar magnitude; \textit{KIC} 8110811 (\textit{TIC} 272270985, \textit{Gaia} DR3 2079577161336956416). \textit{KIC} 8110811 has been observed in sectors 14, 15, 41, 54, 55, 74, 75, 81, and 82 in cadences of 30 minutes, 10 minutes, and 2 minutes. Fortuitously, two of these sectors (54 and 81) overlap with the TESS observations of \bhb and so should have similar, but not identical, noise properties. We calculate a TESS light curve of this target using identical methods as discussed in Sec.~\ref{sec:method} across these overlapping sectors. We find that the signal is observable at the predicted amplitude given Eq.~\ref{eq:ampl}. The noise level of this twin light curve is slightly lower than that of \bhb, 9.3 ppm vs. 10.7 ppm, so it is more likely that the predicted amplitude is incorrect for metal-poor stars.

Of course, it is impossible to prove the absence of a signal despite our best efforts. There are other factors that could hinder detection of the expected 100~$\mu$Hz power excess. The most obvious of these is contamination. Gaia BH3 lies at such a low Galactic latitude ($-3.5^\circ$) that there are several nearby stars, albeit at fainter magnitude. Given the 21" pixel scale of TESS, it is possible that the signal is diluted by companion stars, which we now investigate.

In the 20x20 target pixel file, there are 107 targets brighter than 16th magnitude, but only 6 targets brighter than 13th magnitude (including \bhb). Within the aperture we selected for producing the light curve there is only one other target at 14th magnitude. This target, and the other fainter targets, if they are red giants, have effectively zero probability of showing oscillations in the TESS light curve due to their relative faintness. 

The closest star of similar magnitude to \bhb is \textit{Gaia DR3} 4318465032060792064, at a distance of 56.27" (approximately 3 TESS pixels). This stars color and luminosity imply that it belongs to a more evolved region of the red giant branch. Given the stellar properties from the Gaia DR3 catalog ($\Teff=4905 K$, $\logg=2.5\ \mathrm{dex}$), the star would be expected to be variable at a much lower frequency of $\sim40\ \mu$Hz than what we expect in \bhb.

\subsection{Stellar models}
\label{sec:bhb_model}

We computed a grid of stellar models using Modules for Experiments in Stellar Astrophysics \citep[\texttt{MESA}, version r24.03.1;][]{mesa2011,mesa2013,mesa2015,mesa2018,mesa2019,mesa2023} \footnote{The inlist is available at Zenodo: \dataset[doi: 10.5281/zenodo.16945715
]{\doi{10.5281/zenodo.16945715
}}}. Adiabatic oscillation frequencies were calculated with \texttt{GYRE} \citep[version 7.1;][]{Townsend2013GYRE} using the structure profiles from \texttt{MESA}. The models incorporate elemental diffusion, an Eddington-gray atmospheric boundary condition \citep{Eddington1926}, and exponential convective overshoot at convective envelope boundaries, and a moderate amount of mass loss $\eta = 0.2$ following Reimers' prescription \citep{Reimers1975}. The relative metal abundances follow the solar composition of \cite{Asplund2009Chemical}, with additional $\alpha$-element enrichment to match the observed value of $[\alpha/\mathrm{Fe}]=0.4$.

The initial helium abundance is set to $Y=0.248$, consistent with the primordial value \citep{Planck2016}. Large frequency separations ($\Delta\nu$) were computed using radial modes. 
We sampled the parameter space uniformly over mass $M \in (0.6, 1.2) M_{\odot}$, metallicity $\feh \in (-3.2,-2.4)$, and the mixing length parameter $\alpha_{\textrm{MLT}} \in (1.5,2.5)$.

Gaia BH3 was recently discovered to belong to the metal-poor disrupted ED-2 stellar stream \citep{Dodd2023Gaia, Balbinot2023ED2, Balbinot202433}, a substructure that forms a dynamically cold stellar stream which crosses the solar neighborhood. The near-zero spread in metallicity in ED-2 indicates that it was originated by a disrupted star cluster. This argument is in good agreement with the color-magnitude diagram (cf. fig 1 of \citealt{Balbinot2023ED2}), which is well fitted with an extremely old single stellar population. The CMD of ED-2 closely resembles that of the GC M92 at an age of $13.80\pm0.75$ Gyr \citep{Ying2023Absolute} with a slightly fainter main-sequence turn-off, indicating that \bhb is potentially older than 13 Gyr. This is consistent with the stellar modeling performed in the discovery paper \citep{GaiaCollaboration2024Discovery}, where the de-reddened Gaia colors of \bhb were fit to isochrones of 12 and 14 Gyr. 

\rev{From the stellar models, it is obvious that the mass (along with other stellar parameters) is heavily constrained by the assumed age of the star (Fig.~\ref{fig:model}). Under only metallicity, luminosity, and temperature constraints, the mass of the star is not constrained (as expected). Given that the amplitude of the oscillation signal depends on the mass, the higher mass implies a lower amplitude signal for the same \numax. When using the age constraint of $13\pm1$ Gyr, the parameters are tightly constrained, similar to what was found in \citet{GaiaCollaboration2024Discovery}. Most interestingly however, the predicted amplitude is constrained to a narrow region corresponding to $38\pm3$ ppm (Eq.~\ref{eq:ampl}). This is significantly higher than the median noise level of the power spectrum in the range corresponding to the expected power excess (12 ppm). We also used eq.~19 of \citet{Corsaro2013Bayesian}, who provide a separate scaling relation for the amplitudes. Using this, and the stellar models, we predict an even higher amplitude of $50\pm6$ ppm. Given that we do not detect a signal at either of these levels, it is possible that uncertainty on the luminosity is underestimated, or some other stellar parameter from \citet{GaiaCollaboration2024Discovery} is incorrect. The small uncertainty on the luminosity was likely propagated from their unrealistically small uncertainties on, for example, their adopted reddening value ($A_0 = 0.71\pm0.07$ mag). BH3* exists in a region of low Galactic latitude and is thus subject to a high extinction gradient. The result of these uncertainties means that BH3* barely intersects with the suggested model isochrones within one standard deviation (fig D.2. of \citealt{GaiaCollaboration2024Discovery}). Their photometric temperature is also more than one sigma discrepant to their measured temperature from Fe I lines, casting further doubt on their reported stellar properties.}

Finally, we consider one more scenario for why the predicted signal is not found. The predicted amplitude (Eq.~\ref{eq:ampl}) is scaled from the Kepler data, of which there are only a handful of extremely metal-poor stars. It has also been shown that metallicity can weakly affect the observed mode amplitudes (cf. figure 12 of \citealt{Yu2018Asteroseismology}). It is possible, then, that the metal-poor nature of \bhb has caused us to substantially over-predict its oscillation amplitudes.

Regardless, even a detection of only \numax\ and \dnu\ for \bhb may not be particularly useful for obtaining stellar parameters. As \citet{Huber2024Stellar} note, the asteroseismic scaling relations seem to break down in accuracy for very metal-poor ($\feh<-1$) stars. While constraints on stellar parameters by individual mode-frequency analysis and seismic modeling do remain extremely precise in these situations (e.g. \citealt{Chaplin2020Age,Huber2024Stellar}), this would only be achievable if an improved signal to noise ratio at either a shorter cadence, longer time baseline, or both, were to enable the measurement of individual mode frequencies. Given that we have not detected even a power excess (let alone individual modes), and that \bhb will not fall on any TESS silicon for the foreseeable future \citep{2020ascl.soft03001B}, there is little chance of performing such an analysis without dedicated follow-up. 

\section{Conclusion}

We have investigated the TESS and ground-based light curves of the luminous companions in Gaia BH2 and BH3 for their expected signatures of seismic variability and potential rotational modulation. Our main conclusions are as follows.

\begin{itemize}
    \item We identify an asteroseismic signal in Gaia BH2$^*$ with \numax\ $=$\bhanumax\ and a \dnu\ of \bhadnu. These global asteroseismic parameters yield a mass of $1.19^{+0.08}_{-0.08}$ M$_\odot$, which is in excellent agreement with \citet{El-Badry2023Red}. With only a few TESS sectors of data, we are unable to perform detailed mode identification and modeling.
    \item The asteroseismic age measured for \bha is \bhaage, placing it as a member of the young, $\alpha$-enhanced red giants. This population is suspected to be formed from merger or accretion events, enriching their $\alpha$ elements and increasing their mass.
    \item The long-term ASAS-SN light curve of \bha shows three strong signals of periodicity at 397 d, 495 d, and 1513 days. The ATLAS light curve also shows evidence of photometric modulation at the 397 day period, although longer periods can not be determined given the shorter time coverage of ATLAS. If the signal at 397 d is due to rotational modulation, then it implies that \bha was spun up through some tidal forcing mechanism. That is, the rotation period is faster than permitted by magnetic braking is \bha's rotation period were to be attributed entirely to its primordial angular momentum. We caution that this rotation period is tentative, and requires more follow-up to confirm.
    \item The pseudo-synchronous period inferred from the orbit of \bha is P$_{\text spin} = 428\pm1$ days, which is close to the measured period in the ground-based photometry. This indicates that \bha was potentially spun up in its evolutionary history.
    \item \bha will be observed by TESS in several more sectors over the next few years, after which point individual mode identification and detailed seismic analysis should be possible. This modeling will only serve to improve the age and should indicate whether \bha is indeed a merger product.
    \item We detect no signature of asteroseismic oscillations in \bhb, despite strong constraints on the expected frequency and amplitude of variability. Given the noise level in the light curve, we predict that the expected amplitude should make this variability obvious.
    \item Given the absence of this signal, we compute a series of stellar evolutionary models based on the parameters from \citet{GaiaCollaboration2024Discovery}. We suggest that either their extremely tight luminosity constraint should be relaxed to allow for a larger range of possible amplitudes and frequencies of oscillation, or that the amplitude scaling relations are not well constrained at low metallicity. Our results do not change the inferred mass of the (BH3) black hole companion.
\end{itemize}

Our work demonstrates the utility of asteroseismology to characterize stellar properties, or to confirm the stellar properties obtained from other methods. The TESS mission has now observed a significant fraction of the entire sky, and the ease of producing light curves allows for asteroseismic analyses of brighter objects. Our key takeaway is that the red giant companion to Gaia BH2 is a seemingly young, $\alpha$-enhanced star, with rapid rotation. Both of these phenomena point to the red giant undergoing a merger or accretion event in its past history, which could help constrain theories of dormant black hole binary formation mechanisms.

\section*{acknowledgements}

We thank Tim Bedding, Daniel Huber, Christina Hedges, Kareem El-Badry, and Joel Zinn for helpful comments and suggestions. We also thank the referee for their comments which greatly improved the quality of this manuscript.
    
D. Hey acknowledge support from NSF (AST-2009828), NASA (80NSSC24K0621), and TESS GI Cycle 6 (80NSSC24K0506). Y. Li acknowledges support from Beatrice Watson Parrent Fellowship. J. M. J. Ong acknowledges support from NASA through the NASA Hubble Fellowship grant HST-HF2-51517.001, awarded by STScI. STScI is operated by the Association of Universities for Research in Astronomy, Incorporated, under NASA contract NAS5-26555.

This work has made use of data from the Asteroid Terrestrial-impact Last Alert System (ATLAS) project. The Asteroid Terrestrial-impact Last Alert System (ATLAS) project is primarily funded to search for near earth asteroids through NASA grants NN12AR55G, 80NSSC18K0284, and 80NSSC18K1575.

This paper includes data collected with the TESS mission, obtained from the MAST data archive at the Space Telescope Science Institute (STScI). Funding for the TESS mission is provided by the NASA Explorer Program. STScI is operated by the Association of Universities for Research in Astronomy, Inc., under NASA contract NAS 5–26555.

The playlist made while writing this paper is available \href{https://open.spotify.com/playlist/4ltC2enJSYp3xDQALtD6Hw}{here}.


\vspace{5mm}
\facilities{TESS, ASAS-SN, ATLAS}

\software{astropy \citep{astropycollaborationAstropyCommunityPython2013},  
Modules for Experiments in Stellar Astrophysics
MESA \citep{mesa2011,mesa2013,mesa2015,mesa2018,mesa2019,mesa2023},
TESS-point \citep{2020ascl.soft03001B}.
          }

\begin{singlespace}
    \bibliographystyle{aasjournal}
    \bibliography{library, project}
\end{singlespace}

\end{document}